\documentclass[11pt]{article}
\usepackage{amsmath}
\usepackage{amsfonts}
\usepackage{amssymb}
\usepackage{graphicx} 
\usepackage{times}
\def\figskip{\vskip .5cm plus 3mm minus 2mm}
\begin{document}
%
%
\begin{titlepage}
June 2006 \hfill 
\vskip 5.5cm
{\baselineskip 20pt
\begin{center}
{\bf MIXING ANGLES AND NON-DEGENERATE COUPLED SYSTEMS OF PARTICLES}
\end{center}
}
\vskip .2cm
\centerline{Q. Duret
    \footnote[1]{E-mail: duret@lpthe.jussieu.fr}
          \& B. Machet
     \footnote[2]{E-mail: machet@lpthe.jussieu.fr}
     \footnote[3]{Member of `Centre National de la Recherche Scientifique'}
     }
\vskip 5mm
\centerline{{\em Laboratoire de Physique Th\'eorique et Hautes \'Energies}
     \footnote[3]{LPTHE tour 24-25, 5\raise 3pt \hbox{\tiny \`eme} \'etage,
          Universit\'e P. et M. Curie, BP 126, 4 place Jussieu,
          F-75252 Paris Cedex 05 (France)}
}
\centerline{\em Unit\'e Mixte de Recherche UMR 7589}
\centerline{\em Universit\'e Pierre et Marie Curie-Paris6; CNRS;
Universit\'e Denis Diderot-Paris7}
\vskip 1.5cm
{\bf Abstract:} 
Defining, in the framework of quantum field theory, their mass eigenstates
through their matricial propagator, we show why the mixing matrices
of  non-degenerate coupled systems should not be parametrized as unitary.
This is how, for leptonic binary systems, two-angles solutions  with
discrete values $\pi/4\!\!\mod\!\pi/2$ and $\pi/2\!\!\mod\!\pi$ arise
when  weak leptonic currents of mass eigenstates approximately
satisfy the two properties of universality and vanishing of their
non-diagonal neutral components. Charged weak currents are also discussed,
which leads to a few remarks concerning oscillations.
We argue that quarks, which cannot be defined on shell because of the
confinement property,
are instead more naturally endowed with unitary Cabibbo-like mixing
matrices, involving a single unconstrained mixing angle.
The  similarity  between neutrinos and neutral kaons is outlined, together
with the role of the symmetry by exchange of families.

\bigskip

{\bf PACS:} 11.10.-z\quad 12.15.Ff\quad 14.60.Pq
\vfill
\begin{center}
\includegraphics[height=1.8truecm,width=4truecm]{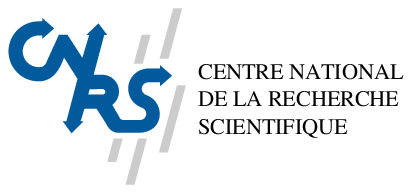}
\end{center}
\end{titlepage}
%

\section{Introduction}
\label{section:introduction}

The observed large mixing angles in the neutrino sector have already long
been a matter of surprise and questioning \cite{Kayser}\cite{AFM}.
Symmetries \cite{Ma} have been invoked, which should be  approximate since
the mixing is only close to maximal, various ``textures'' of mass
matrices have been proposed \cite{Alt}, \cite{Smir}
which are not fully satisfying either, and furthermore unstable
by  unitary transformations on flavour eigenstates \cite{Jarlskog1}.

In the while, the $CP$ violating parameters $\epsilon_L$ and
$\epsilon_S$ of the physical neutral kaons $K_L$ and $K_S$ have been shown
\cite{MaNoVi}\cite{Novikov}, using the propagator formalism
of quantum field theory, not to be rigorously identical. This
amounts to a tiny lack of unitarity in the mixing matrices linking ``in'' (or
``out'') mass eigenstates to the orthonormal basis of flavour eigenstates.
Phrased in another way, mixing matrices of physical kaons cannot be
parametrized with a single mixing angle. It was also shown that
systems of this type should not be described by a single constant
mass matrix and that  their mass eigenstates cannot be correctly determined
by a bi-unitary transformation.

In this  note we establish a link between these two peculiarities 
and show they are among  general properties of non-degenerate coupled
systems of particles.

\section{General framework}
\label{section:general}

\subsection{Flavour and mass eigenstates}

Since their couplings to the Higgs boson are not flavor-diagonal,
massive fermions in the standard model form  coupled systems
(like neutral kaons).
The usual approach to such systems makes use of a mass matrix (see appendix
\ref{section:mama}).
It was however shown in  \cite{MaNoVi}\cite{Novikov}
that it is an inadequate procedure: indeed, a (constant) mass matrix
can  only be introduced as  a linear
approximation to the inverse propagator in the vicinity of
 each of its poles, such that, in the case under study, 
as many mass matrices as there are poles should be considered.
This is why we  stick below to basic principles
of quantum field theory, which state in particular that the physical masses
of particles, bosons or fermions,
can only be the poles of their full propagator. The corresponding
eigenstates -- the propagating states -- are the mass / spin
eigenstates of the Lorentz-Poincar\'e group.

Two bases play a fundamental role, in particular in the electroweak physics
of leptons: the $2n_f$ flavour eigenstates
\footnote{$n_f$ denotes here the number of families.}
 ($e^-_f, \mu^-_f, \nu_{e,f},
\nu_{\mu,f} \ldots$) which, by convention, couple to weak vector bosons, 
and  the $2n_f$ propagating eigenstates
($e^-_m, \mu^-_m, \nu_{e,m}, \nu_{\mu,m} \ldots$), which are also the
mass eigenstates. At the classical level only left-handed flavour
 fermions weakly couple, but the situation changes when quantum corrections
are included.

The {\em physical masses} $z_i=m_i^2$ satisfy by definition
the gauge invariant pole equation (the variable $z$ is used for $q^2$)
\begin{equation}
\det \Delta^{-1}(z) =0,\ \textrm{for}\  z=z_i,
\label{eq:pole1}
\end{equation}
where $\Delta(z)$ is the  full (renormalized) $2n_f \times 2n_f$ matrix
propagator in momentum space.
The solutions of (\ref{eq:pole1}) are independent of the renormalization
procedure.
The propagating (mass) eigenstates $\varphi^i_{m}$  are the
corresponding eigenvectors  with vanishing eigenvalues
\begin{equation}
\Delta^{-1}(z=z_i) \varphi^i_{m} = 0.
\label{eq:pole2}
\end{equation}
It is also convenient to introduce $\Delta^{-1}(z) = L^{(2)}(z)$ as
the renormalized quadratic Lagrangian operator.
(\ref{eq:pole1}) then reads
\begin{equation}
\det L^{(2)}(z)=0,
\label{eq:L1}
\end{equation}
and the mass eigenstates
satisfy  the equation (equivalent to (\ref{eq:pole2}))
\begin{equation}
L^{(2)}(z=z_i) \varphi^i_{m} =0.
\label{eq:L2}
\end{equation}
The situation is accordingly that of a $z$ dependent $2n_f\times 2n_f$ matrix
$L^{(2)}(z)$, the $2n_f$ eigenvalues $\lambda_j(z), j = 1 \ldots 2n_f$ of
which are supposed non-degenerate and satisfy, by definition of the poles
$z_i$ of $\Delta(z)$, $\lambda_i(z_i) = 0$.
At any $z$ it has $2n_f$ eigenvectors $\psi^j(z), j=1 \ldots 2n_f$.
When $z \to z_i$, $\psi^i(z) \to \varphi^i_m$ and $\psi^j(z), j\not=i \to
\omega^j_i$. So, among the $2n_f$ eigenvectors of $L^{(2)}(z_i)$
 lies the mass eigenstate $\varphi^i_m$ corresponding to the vanishing
eigenvalue and $2n_f-1$ other eigenstates
$\omega^j_{i\not=j}$, that we call spurious \cite{MaNoVi}\cite{Novikov},
and which correspond
to non-vanishing eigenvalues $\lambda_j(z_i), j\not=i$. They just
represent off-mass-shell states. The case of two flavors ($n_f=1$)
 is depicted on Fig.~1 below.

Mixing matrices link flavour ($\Psi_f$) to mass ($\Psi_m$) eigenstates:
 simplifying to $n_f=2$ ($4$ flavors)
\begin{equation}
\Psi_f = J \Psi_m,\quad
\Psi_f = \left(\begin{array}{c} \nu_{e,f} \cr \nu_{\mu,f} \cr e^-_f \cr
\mu^-_f\end{array}\right),\quad
\Psi_m = \left(\begin{array}{c} \nu_{e,m} \cr \nu_{\mu,m} \cr e^-_m \cr
\mu^-_m\end{array}\right),\quad
J= \left(\begin{array}{ccc} K_\nu & \vline & \cr\hline
                                      & \vline & K_{\ell} \end{array}\right),
\label{eq:mixmat}
\end{equation}
where we have split every $2n_f \times 2n_f$ matrix into four $n_f \times
n_f$ sub-blocks.
The entries of $\Psi_m$ are the $\varphi_m^i$'s of (\ref{eq:pole2}) and
(\ref{eq:L2}).

The (renormalized) quadratic Lagrangian density is ${\cal L}^{(2)}(x) =
\overline{\Psi_f} L^{(2)}(x) \Psi_f$.  

Fermions are usually considered as bi-spinors (of Dirac or Majorana types)
built from two Weyl spinors of different
chiralities  which are also orthogonal to each other.
Two sets of different mixing matrices therefore generally occur,
respectively for left and right spinors (see also appendix \ref{section:mama}).
$K_\nu$ and $K_\ell$ in (\ref{eq:mixmat}) must then be also attributed a
subscript $L$ or $R$ depending on which chirality  is considered.
In order not to  overload the notations, these subscripts  will be
understood in the following.

\subsection{Mixing matrices for mass-split on-shell fermions are not
unitary}

As already mentioned in \cite{MaNoVi}\cite{Novikov}, the connection
between flavour eigenstates and
non-degenerate mass eigenstates is not a unitary transformation. Indeed:
\newline
in flavour space, $L^{(2)}(z)$ being, at each $z$, a hermitian $2n_f \times
2n_f$ operator (matrix), its $2n_f$ eigenstates form an orthonormal basis
$\Psi(z)$ (because it is in particular normal, left and right eigenstates
coincide).
At $z = z_i= m_i^2$,
$<\varphi^i_m\ |\ \varphi^i_m>=1$ and
$<\varphi^i_m\ |\ \omega^j_i, j\not=i>=0$.\quad
Thus, at the $2n_f$ values $z=z_i$, $2n_f$ different orthonormal bases
(of $2n_f$ eigenstates) occur. Since two non-degenerate mass eigenstates
$\varphi^i_m$ and $\varphi^k_m$ belong to two different orthonormal
bases, they are in general not orthogonal:
\begin{equation}
<\varphi^i_m\ | \ \varphi^k_m> \not = 0, \quad i \not= k.
\label{eq:nonorth}
\end{equation}
This being true in  the  neutral and charged sectors,  both $K_\nu$ and
$K_{\ell}$, which  connect the flavour basis to a non-orthonormal one,
 have no reasons to be unitary
\begin{equation}
K_{\ell}^\dagger K_{\ell} \not = 1, \quad K_\nu^\dagger K_\nu \not=1, \quad q.e.d.
\label{eq:nonunit}
\end{equation}
The non-unitarity of mixing matrices does not however jeopardize the unitarity
of the theory (see appendix \ref{section:unit}).
It simply states that, at a given $q^2$, all physical states
cannot be simultaneously on-shell when they are non-degenerate.

It may happen, for example to describe unstable particles (like neutral
kaons), that one is led to introduce an (effective) Hamiltonian,
or Lagrangian, which is non-hermitian, and even non-normal. Then, at each
$z$, the set of eigenstates $\psi^j(z)$ do not form any more an orthonormal
basis. Spurious states still accompany the mass eigenstate at $z=z_i$.
 Different mass eigenstates, corresponding to
different $z_i$'s, have no reason either in this case
 to form an orthonormal basis, as
explicitly checked in \cite{MaNoVi}.

The simplest case of two flavours ($n_f=1$) is depicted on Fig.~1
which represents either the neutral kaon system, or, in the cases
of two lepton families,  the neutrino sector or the charged lepton sector. 
The $z$-independent flavour basis $(\psi_1, \psi_2)$
(for example $(K^0,\overline{K^0})$ for neutral kaons)
 has been represented by the two horizontal lower lines.

\bigskip

\vbox{
\begin{center}
\includegraphics[height=8truecm,width=10truecm]{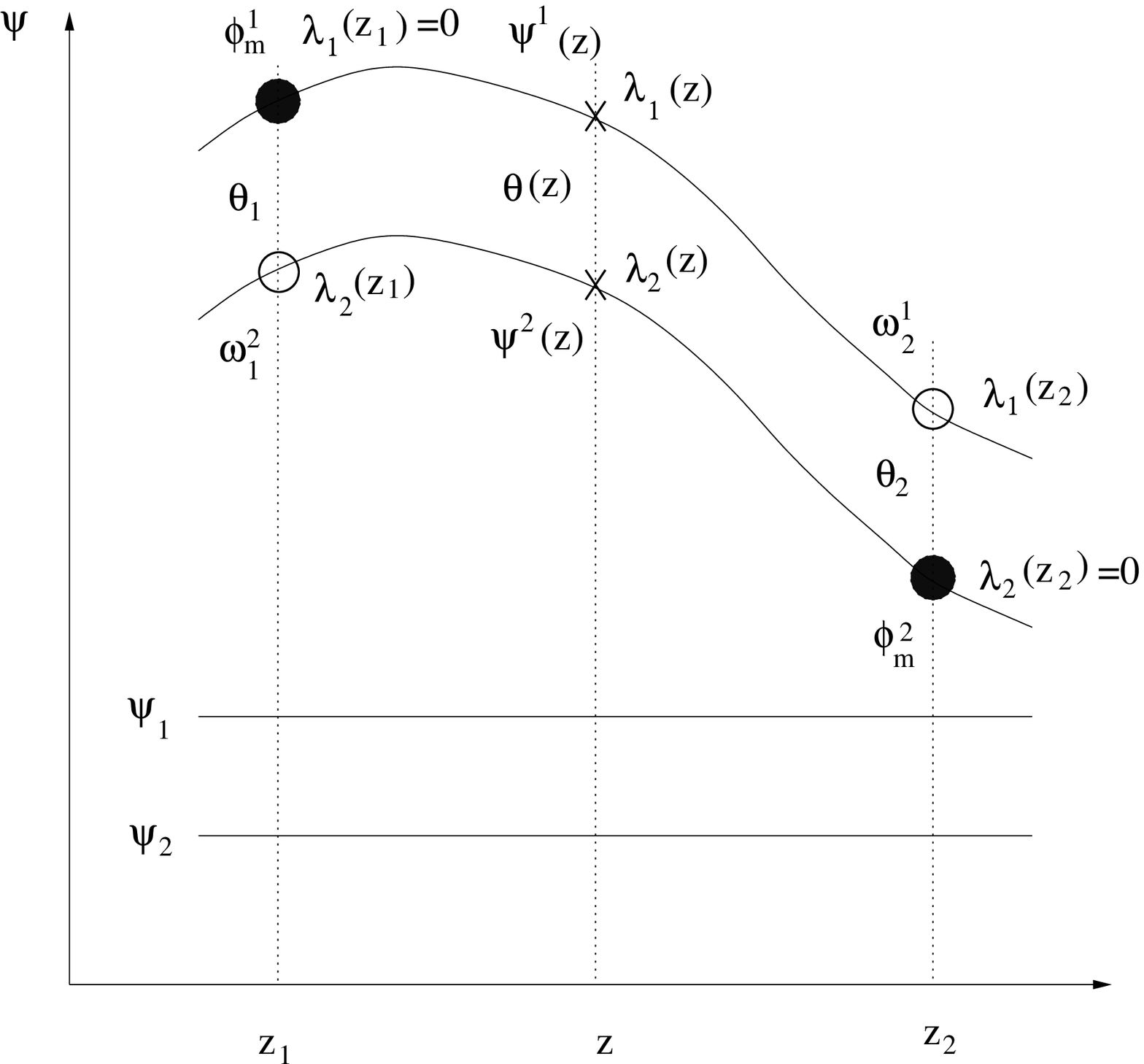}
\end{center}

\centerline{{\em Fig.~1: Eigenstates of a binary coupled system}}
}

\bigskip

The two eigenstates of the
hermitian (normal) renormalized propagator $\psi^1(z)$ and $\psi^2(z)$ form
a $z$-dependent orthonormal basis. When
$z$ varies, this builds up an  infinite set of orthonormal bases which is
depicted by the two (parallel) curved lines.
At a given z, the orthonormal basis $(\psi^1(z), \psi^2(z))$ is connected
to the orthonormal flavour basis by a unitary mixing matrix with angle
$\theta(z)$.
At $z=z_1$, $\varphi^1_m$ and $\omega^2_1$ form an orthonormal basis,
and so do, at $z=z_2$, $\varphi^2_m$ and $\omega^1_2$.
They are respectively related to the basis of flavour eigenstates by two
different unitary matrices, with respective angles $\theta_1$ and
$\theta_2$.
$\varphi^1_m$ and $\varphi^2_m$ do not form in general an orthonormal basis,
and it intuitively appears on the picture that the mixing matrix
connecting them to the flavour basis cannot be parametrized with a single
angle (both angles $\theta_1$ and $\theta_2$ obviously play a role).

\subsection{The case of quarks}

(\ref{eq:nonunit}) applies to states for which the full propagator has
poles, corresponding to physical (``on-shell'') propagating states which
can be identified with particles.
In contrast, quarks are never produced on shell: the poles of their full
propagator are ill-defined and so are accordingly their ``physical'' masses
and mass splittings.
The only unambiguous orthonormal basis which then occurs in $L^{(2)}$
(supposed to be hermitian) is the $z$ dependent basis $\psi^j(z)$.
At each $z$ are associated two unitary, $z$-dependent mixing matrices
$K_u(z)$ and $K_d(z)$. Their unitary product $K(z) = K_u^\dagger(z) K_d(z)$
we propose to consider as the equivalent of the renormalized unitary
Cabibbo-Kobayashi-Maskawa (CKM) matrix of the standard approach,
 in which the complex mass matrices
$M_u$ and $M_d$ generated by the couplings of quarks to the Higgs boson are
diagonalized by bi-unitary transformations  (see also appendix
\ref{section:mama}).

\section{Leptonic weak currents}

\subsection{Fermion coupling to weak gauge bosons}

In the flavour basis $\Psi_f$, the weak Lagrangian reads

\vbox{
\begin{eqnarray}
&&{\cal L}_{weak}
=\overline{\Psi}_f\gamma^\mu\frac{1-\gamma^5}{2}\Big[W_\mu^+ T^+ + W_\mu^- T^- + W_\mu^3
T^3\Big]\Psi_f,\cr
&& \cr
&& T^+ = \left(\begin{array}{ccc} & \vline & 1 \cr \hline & \vline &
\end{array}\right),
T^- = \left(\begin{array}{ccc} & \vline &  \cr \hline 1 & \vline &
\end{array}\right),
T^3 = \left(\begin{array}{ccc} 1 & \vline & \cr \hline & \vline & -1 
\end{array}\right).
\label{eq:Lf}
\end{eqnarray}
}

$T^\pm, T^3$ form a representation of the $SU(2)$ group of weak
interactions: $[T^+,T^-] = T^3$,  etc.
In the orthonormal basis $\Psi(z)$ one finds another $SU(2)$ representation
$[\widehat T^+(z),\widehat T^-(z)]=\widehat T^3(z)$:
\begin{eqnarray}
&&{\cal L}_{weak}
=\overline{\Psi}(z)\gamma^\mu\frac{1-\gamma^5}{2}\Big[W_\mu^+ \widehat T^+(z) + W_\mu^- \widehat T^-(z) + W_\mu^3
\widehat T^3(z)\Big]\Psi(z),\cr
&& \cr
&&\hskip -1cm \widehat T^+ = \left(\begin{array}{ccc} & \vline & K^\dagger_\nu(z)
K_\ell(z) \cr \hline & \vline &
\end{array}\right),
\widehat T^- = \left(\begin{array}{ccc} & \vline &  \cr \hline 
 K_\ell^\dagger(z) K_\nu(z) & \vline &
\end{array}\right),
\widehat T^3 = \left(\begin{array}{ccc} 1 & \vline & \cr \hline & \vline & -1 
\end{array}\right).
\label{eq:Lz}
\end{eqnarray}
In the non-orthonormal basis $\Psi_m$ of mass eigenstates,
(\ref{eq:Lf}) becomes
\begin{eqnarray}
&&{\cal L}_{weak}=\overline{\Psi}_m
\gamma^\mu\frac{1-\gamma^5}{2}\Big[W_\mu^+ {\mathfrak T}^+ + W_\mu^- {\mathfrak T}^- + W_\mu^3
{\mathfrak
T}^3\Big] \Psi_m, \quad {\mathfrak T}^i = J^\dagger T^i J,\cr
&& \cr
&& \hskip -1cm{\mathfrak T}^+ = \left(\begin{array}{ccc} & \vline &
K_\nu^\dagger K_{\ell} \cr \hline & \vline &
\end{array}\right),
{\mathfrak T}^- = \left(\begin{array}{ccc} & \vline &  \cr
\hline K_{\ell}^\dagger K_\nu & \vline &
\end{array}\right),
{\mathfrak T}^3 = \left(\begin{array}{ccc} K_\nu^\dagger K_\nu
& \vline & \cr \hline & \vline & -K_{\ell}^\dagger K_{\ell} 
\end{array}\right)
\label{eq:Lm}
\end{eqnarray}
and, because of (\ref{eq:nonunit}),
the $SU(2)$ commutation relations are not systematically satisfied
by the ${\mathfrak T}$'s  (and $\Psi_m$ does not simply decompose
into two $SU(2)$ doublets).

$K_\nu$ is related to $K_\nu(z_1)$ and $K_\nu(z_2)$ by
\begin{eqnarray}
K_\nu &=& \frac{1}{K_{\nu[22]}(z_1)K_{\nu[11]}(z_2) - K_{\nu[12]}(z_1)K_{\nu[21]}(z_2)}
\left(\begin{array}{cc} {\cal D}_\nu(z_1) K_{\nu[11]}(z_2) & {\cal D}_\nu(z_2)K_{\nu[12]}(z_1)\cr
{\cal D}_\nu(z_1) K_{\nu[21]}(z_2) & {\cal D}_\nu(z_2)
K_{\nu[22]}(z_1)\end{array}\right),\cr
&& \cr
&& {\cal D}_\nu(z) = \det K_\nu(z),\  z_1=m_{\nu_{e_m}}^2,\  
z_2=m_{\nu_{\mu_m}}^2.
\end{eqnarray}
The following exact relations
\begin{equation}
<\varphi^1_m\ |\ \psi_1>=K_{[11]}(z_1), 
<\varphi^2_m\ |\ \psi_1>=K_{[12]}(z_2), 
<\varphi^1_m\ |\ \psi_2>=K_{[21]}(z_1), 
<\varphi^2_m\ |\ \psi_2>=K_{[22]}(z_2),
\label{eq:scalprod}
\end{equation}
also hold, which become compatible with the approximate formula 
\begin{equation}
 K_\nu \simeq \left(\begin{array}{cc} 
K_{\nu [11]}(z_1) & K_{\nu[12]}(z_2)\cr
 K_{\nu[21]}(z_1) & K_{\nu[22]}(z_2)\end{array}\right)
\label{eq:approxsp}
\end{equation}
when one neglects the scalar products $<\varphi^1_m\ |\ \varphi^2_m>$
supposed to be small. When $z_2 \to z_1$, (\ref{eq:scalprod}) and
(\ref{eq:approxsp}) give back a unitary
mixing matrix $K_\nu(z_1) \equiv \lim_{z_2\to z_1} K_\nu$. This shows
the role of the non-degeneracy in the non-unitarity of $K_\nu$.

\subsection{Weak currents}

It is remarkable, but often unnoticed that, in the quark sector of the
standard model, with unitary $K_u$ and $K_d$
\footnote{$K_u$ and $K_d$ are the mixing matrices respectively
 for $u$-type and $d$-type quarks.}
, the built-in characteristic of
the weak Lagrangian that the couplings of flavour fermions to gauge bosons
are symmetric by the exchange of families,
translates into a similar property for mass states,
without any constraint on the (then unique) mixing angle.
The same holds concerning the absence of flavour changing neutral
currents.

The situation becomes different for  mass-split physical states.
Considering the case of two families, when a single mixing angle is
not enough to describe the system, 
the symmetry by exchange  of families for mass states (that we  call
universality for mass states) and the absence of ``mass changing
neutral currents'' (MCNC's) are no longer automatically
achieved. They instead  require well defined relations between  mixing angles.
We demonstrate below that, within each $2 \times 2$
subspace of given electric charge,  they constrain the two
mixing angles to be: either $\pi/2 \mod\pi$, 
or $\pi/4 \mod\pi/2$. In the first case
the family indices of flavour and mass states are either identical or
crossed. The last case corresponds to the ``maximal mixing''
(approximately) observed for neutrinos and for neutral kaons.
Mass eigenstates are then symmetric or antisymmetric by family exchange
(for example so are the $PC$ eigenstates $K^0 \pm \overline{K^0}$).
Maximal mixing  accordingly realizes universality in both spaces of flavour
and mass eigenstates.

\subsubsection{Neutral weak currents of mass eigenstates}

We deal with weak  currents of mass eigenstates and investigate
the property that MCNC's are very small and that their diagonal
counterparts are
quasi-universal.
Allowing a lack of unitarity (\ref{eq:nonunit}), we parametrize,
with transparent notations (preserving a unit norm for all states and
discarding irrelevant global phases)
\begin{equation}
K_\nu = \left(\begin{array}{rr} e^{i\alpha}c_1 & e^{i\delta}s_1 \cr
-e^{i\beta}s_2 & e^{i\gamma}c_2
\end{array}\right),\ 
K_{\ell} = \left(\begin{array}{rr} e^{i\theta}c_3 & e^{i\zeta}s_3 \cr
-e^{i\chi}s_4 & e^{i\omega}c_4
\end{array}\right).
\label{eq:Kparam}
\end{equation}
Note that
 $[K_\nu,K^\dagger_\nu] \not=0$: $K_\nu$ and $K_\ell$ are not normal.
(\ref{eq:Kparam})  entails
\begin{eqnarray*}
K_\nu^\dagger K_\nu &=& \left(\begin{array}{cc}
c_1^2 + s_2^2 & c_1 s_1e^{i(\delta -\alpha)} -c_2 s_2e^{i(\gamma-\beta)}
 \cr c_1 s_1e^{i(\alpha -\delta)} -c_2 s_2e^{i(\beta-\gamma)} & s_1^2 + c_2^2
\end{array}\right),\cr
K_{\ell}^\dagger K_{\ell} &=& \left(\begin{array}{cc}
c_3^2 + s_4^2 & c_3 s_3e^{i(\zeta-\theta)} -c_4 s_4e^{i(\omega-\chi)}
\cr c_3 s_3 e^{i(\theta-\zeta)} -c_4 s_4 e^{i(\chi-\omega)} & s_3^2 + c_4^2
\end{array}\right).
\end{eqnarray*}
The two requests of (quasi) universality and absence of MCNC's,
equivalent to $K_\nu^\dagger K_\nu \approx
1 \approx K_\ell^\dagger K_\ell$, translate into, respectively,
the identity  of diagonal elements and the vanishing of non-diagonal elements.
The condition $K_\nu^\dagger K_\nu =1$, in the simple case where
the phases $\alpha,\beta,\gamma,\delta$ are vanishing,  can be visualized
on Fig.~2, which is drawn in the orthonormal flavour basis:
\newline\null
\quad -\ the two unit vectors $(c_1,s_1)$ and $(-s_2,c_2)$ are the two
dotted vectors;\newline\null
\quad -\ the mass eigenstates proportional to $(c_2, -s_1)$ and $(s_2,c_1)$
are the two green vectors;\newline\null
\quad -\ all vectors are uniquely determined by $\theta_1$ and $\theta_2$;
the condition under scrutiny is that of finding these two angles
 such that the red and blue vectors $(c_1,-s_2)$ and
$(s_1,c_2)$ (or, equivalently, the mass eigenstates) become {\em orthonormal}.

\vbox{
\begin{center}
\includegraphics[height=5cm, width= 6cm]{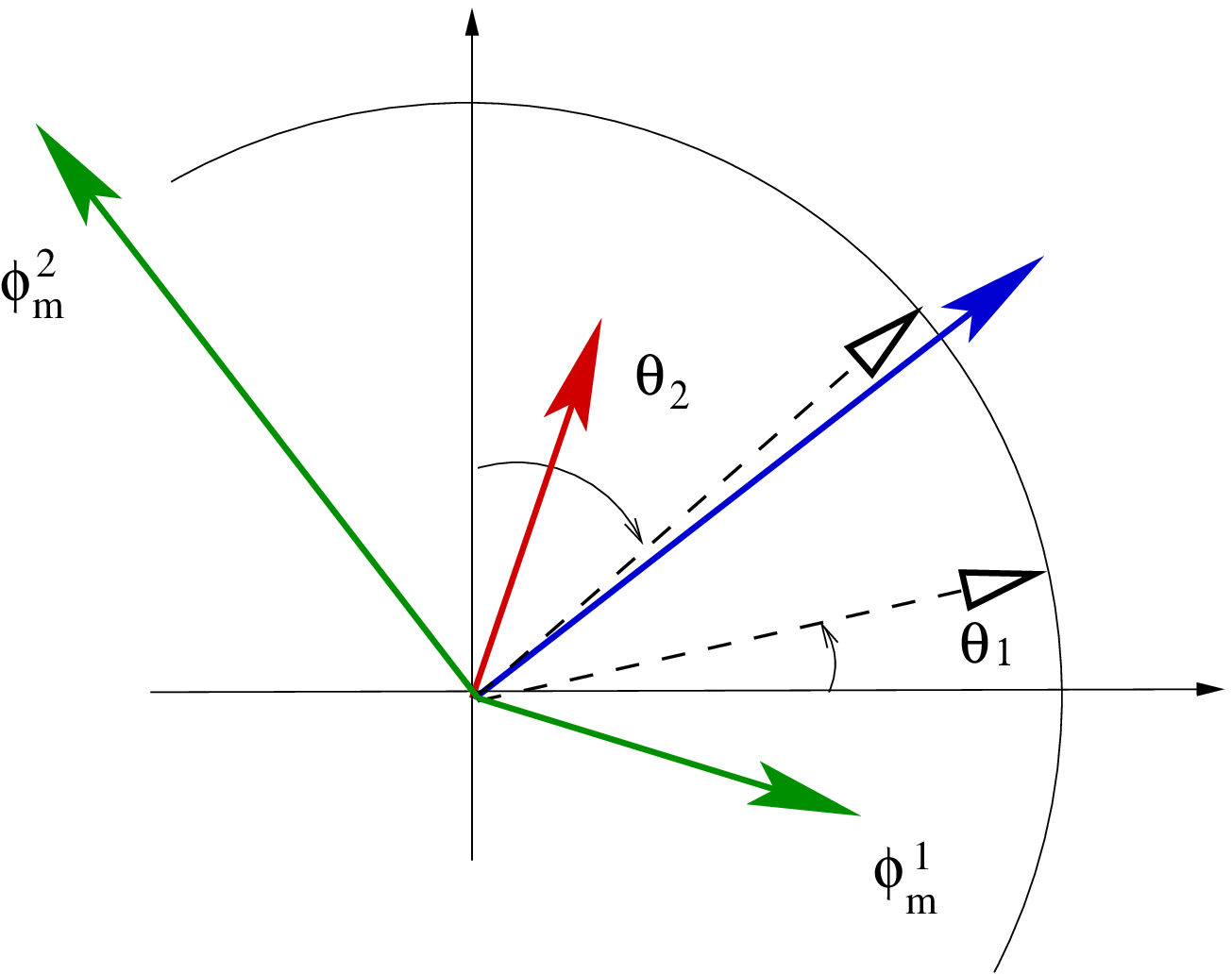}
\hskip 2cm\null
\includegraphics[height=5cm, width= 5.5cm]{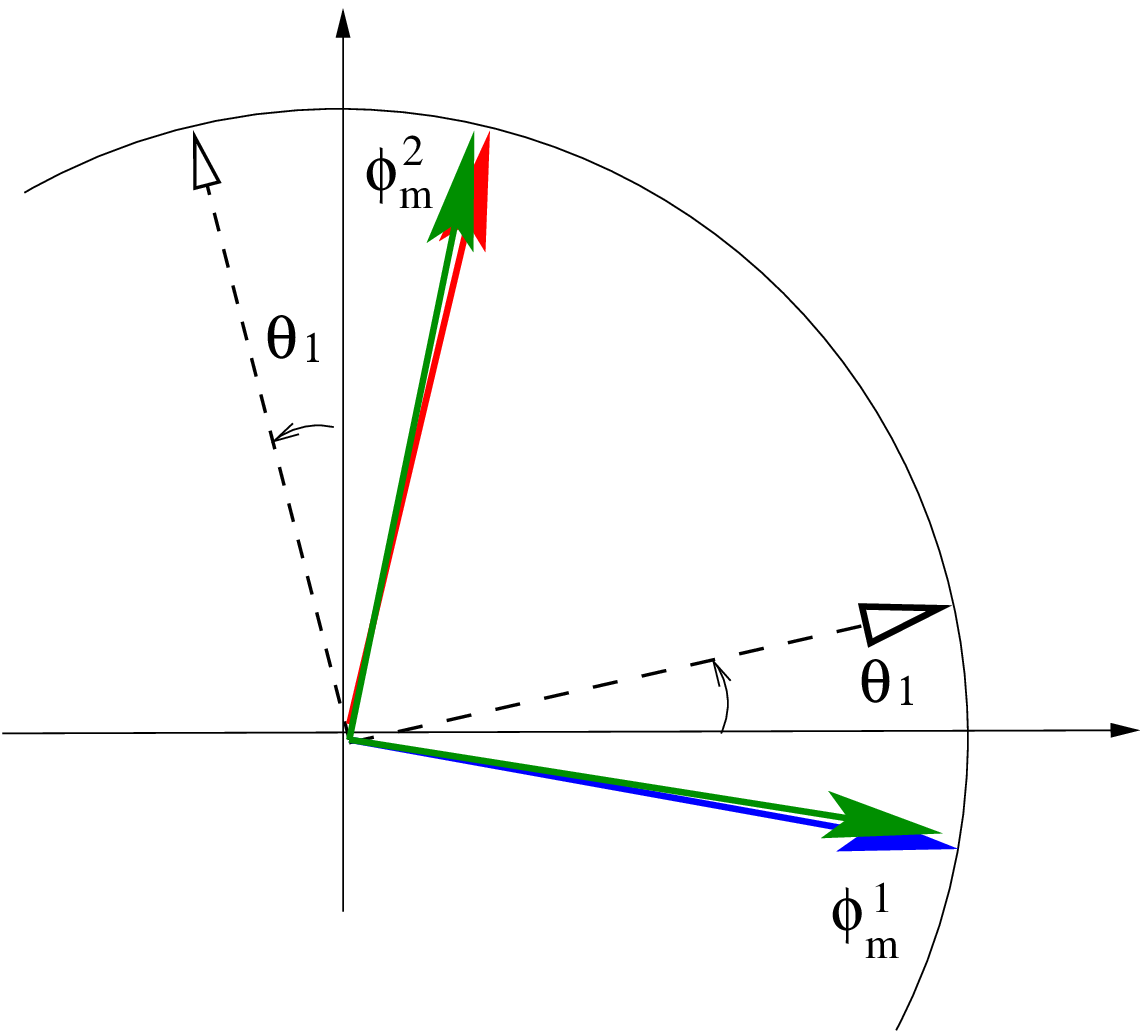}

{\em Fig.~2: graphical representation of the two conditions of universality and
absence of MCNC's;\break on the right is a ``Cabibbo-like'' solution
$\theta_2=\theta_1$.}
\end{center}
}

The discussions for neutrinos and for charged leptons being similar,
we proceed with the former.

$\bullet$\ Quasi-universality is satisfied for $c_1^2 + s_2^2 = c_2^2 + s_1^2
\Leftrightarrow c_1^2 = c_2^2$  which requires
 $\theta_2 \approx \pm \theta_1 + k\pi$ (a).

$\bullet$ The quasi-absence of MCNC's requires

\hskip 2cm $\ast$ either: $c_1s_1 = c_2s_2$ (f\raise-2pt\hbox{\tiny 1})
and $e^{i(\alpha-\delta)} =
e^{i(\beta-\gamma)}$ (f\raise-2pt\hbox{\tiny 2}),\newline\null
\hskip 2cm $\ast$ or: $c_1s_1 = -c_2s_2$ (g\raise-2pt\hbox{\tiny 1}) and
$e^{i(\alpha-\delta)}=-e^{i(\beta-\gamma)}$ (g\raise-2pt\hbox{\tiny 2}).

(f\raise-2pt\hbox{\tiny 1}) requires either
 $\theta_2 \approx \theta_1 + n\pi$ (b), or
$\theta_2  \approx  -\theta_1 + \pi/2 + n\pi$ (c), while
(g\raise-2pt\hbox{\tiny 1}) requires either
$\theta_2 = -\theta_1 + n\pi$ (d) or $\theta_2 = \theta_1 + \pi/2 + n\pi$ (e).
The different cases to consider are accordingly (a) $\cup$ (b) $\cup$ (f\raise-2pt\hbox{\tiny 2}),
(a) $\cup$ (c) $\cup$ (f\raise-2pt\hbox{\tiny 2}), (a) $\cup$ (d) $\cup$
(g\raise-2pt\hbox{\tiny 2}) and (a) $\cup$ (e)
$\cup$ (g\raise-2pt\hbox{\tiny 2}).\newline\null
The solutions of (a) $\cup$ (b) $\cup$ (f\raise-2pt\hbox{\tiny 2}) and (a)
$\cup$ (d) $\cup$ (g\raise-2pt\hbox{\tiny 2}) are

\vbox{
\begin{eqnarray}
(a)\cup (b) \cup (f\raise-2pt\hbox{\tiny 2})&:& \theta_2 = \theta_1 + k\pi\
\textrm{or}\ 
\big\{\theta_1 = (k-n)\frac{\pi}{2}, \theta_2 = (k+n)\frac{\pi}{2}
= -\theta_1 + k\pi\big\},\cr
(a)\cup (d) \cup (g\raise-2pt\hbox{\tiny 2})&:& \theta_2 = -\theta_1 +
k\pi\ \textrm{or}\ 
\big\{\theta_1 = (n-k)\frac{\pi}{2}, \theta_2 = (n+k)\frac{\pi}{2}
= \theta_1 + k\pi\big\}.
\end{eqnarray}
}

The solutions of (a) $\cup$ (c) $\cup$ (f\raise-2pt\hbox{\tiny 2}) and (a)
$\cup$ (e) $\cup$ (g\raise-2pt\hbox{\tiny 2}) are
\begin{eqnarray}
(a)\cup (c) \cup (f\raise-2pt\hbox{\tiny 2})&:& 
\big\{\theta_1 = \frac{\pi}{4} + (n-k)\frac{\pi}{2}, \theta_2 =
\frac{\pi}{4} + (n+k)\frac{\pi}{2}=\theta_1 + k\pi\big\},\cr
(a)\cup (e) \cup (g\raise-2pt\hbox{\tiny 2})&:& 
\big\{\theta_1 = -\frac{\pi}{4} + (k-n)\frac{\pi}{2}, \theta_2 =
\frac{\pi}{4} + (k+ n)\frac{\pi}{2}=-\theta_1+k\pi\big\}.
\end{eqnarray}
There exist accordingly two sets of solutions:

\quad$\ast$\ Cabibbo-like solutions $\theta_2 = \pm \theta_1 + k\pi$
 for which the equations for universality
and for the absence of MCNC's coincide: the $(\theta_1,\theta_2)$
surfaces defined by $c_1^2 = c_2^2$ and $c_1s_1=c_2s_2$ intersect along a
line, which yields a one parameter solution (with a single, unconstrained,
 mixing angle).

\quad$\ast$\  Cases for which the two equations are  independent: the
  two surfaces (one of which  can be checked to always present a saddle
point in the vicinity of the intersections) intersect at  discrete points.

\vbox{
\begin{center}

\includegraphics[height=8cm, width= 10cm]{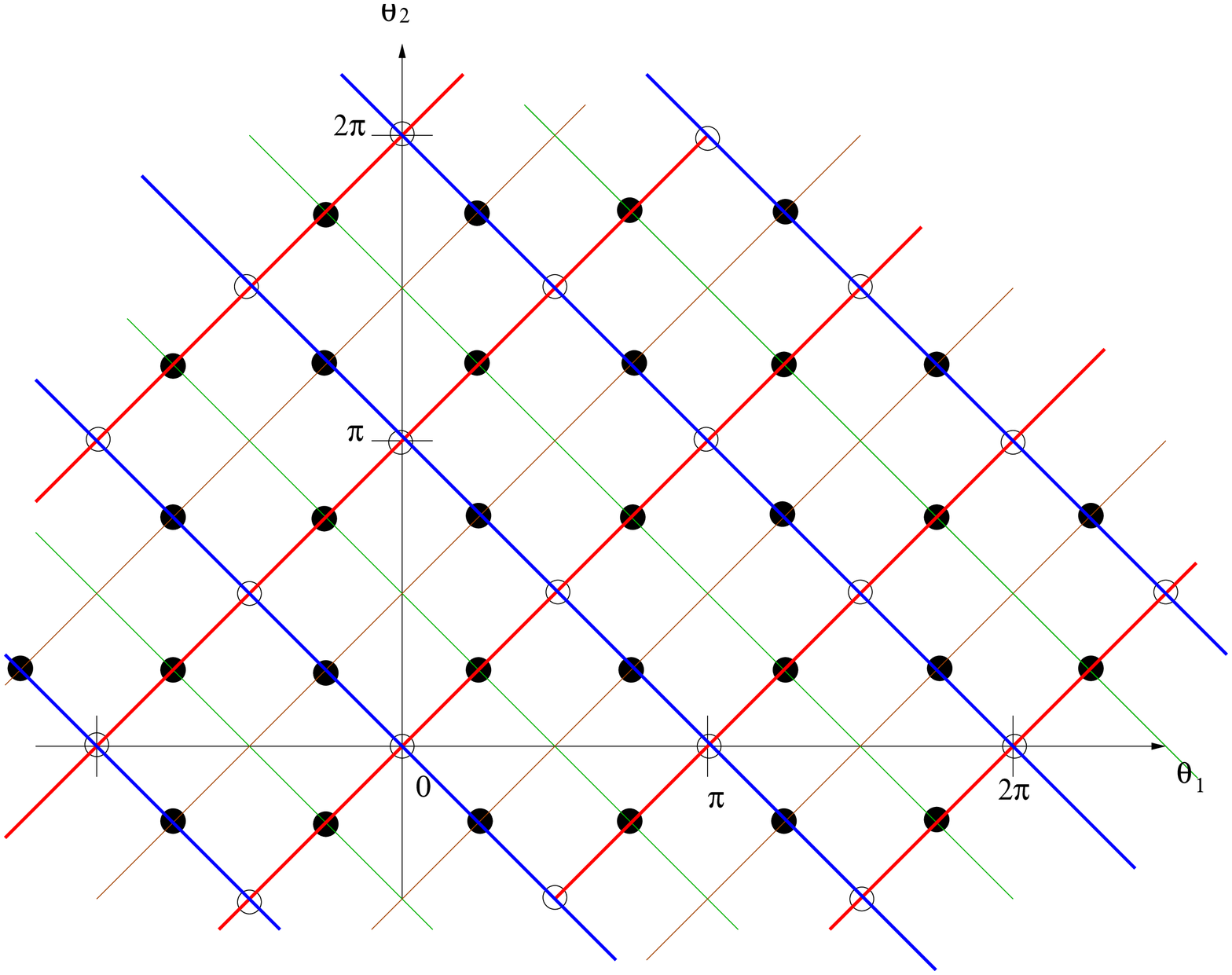}

{\em Fig.~3: graphical solution of the two conditions of universality and
absence of MCNC}
\end{center}
}

The set of all solutions is depicted on Fig.~3 (in which 
the conditions (f\raise-2pt\hbox{\tiny 2}) or (g\raise-2pt\hbox{\tiny 2})
are supposed to be realized).
It is made of the entire red and blue lines $+$ all black and white dots.
The blue and red lines correspond respectively to $\theta_2 = \theta_1  +
k\pi$ and $\theta_2 = -\theta_1 + k\pi$ (conditions (a), (b) or (d)). They
represent the Cabibbo-like situations. They cross (white dots)
 at the discrete values $\pi/2 +k\pi$ of $\theta_1$ and $\theta_2$.
The brown and green lines correspond respectively to
$\theta_2 = \pi/2 + \theta_1 + k\pi$ (c) and $\theta_2 = \pi/2 -\theta_1 +
k\pi$ (e). Their intersections (black dots) with the blue
or red lines (on which in particular condition (a) holds) provide the maximal
mixing solutions $\pm \pi/4 + n\pi/2$.\newline\null
Note that $\theta_{1,2} \approx 0, \pi$ are allowed for physical
non-degenerate particles. The case $\theta=0$ means that mass and flavour
eigenstates are exactly aligned (which is usually assumed for charged leptons).

Though discrete solutions are also located on red or blue lines, they
should not be mixed up with Cabibbo-like solutions: the former are
one-parameter solutions,  while the latter depend on two parameters.

Both types, when exact, can be shrunk, by rephasing the fermions,
 to a single mixing angle which is unconstrained
for Cabibbo-like cases and has  fixed values for others.
We give below a Cabibbo-like example
\footnote{Since they  lie on the trajectories of
Cabibbo-like solutions, this is also a general property of all exact
discrete solutions.}
.
For $\theta_2 = \theta_1 + \pi$ and $e^{i(\delta-\alpha)} =
 e^{i(\gamma-\beta)} = e^{i\xi}$ ({\em i.e.} for conditions
(b) $\cup$ (f\raise-2pt\hbox{\tiny 2})), or  $\theta_2 = -\theta_1$
and $e^{i(\delta-\alpha)} = - e^{i(\gamma-\beta)} = e^{i\xi}$ ({\em i.e.}
for conditions (d) $\cup$ (g\raise-2pt\hbox{\tiny 2})),  one has
\begin{equation}
\left(\begin{array}{c} e^{-i\alpha}\nu_{em}\cr e^{-i\beta}
\nu_{\mu m}\end{array}\right)
= \left(\begin{array}{rr} \cos (-\theta_1) & \sin(-\theta_1) \cr
-\sin(-\theta_1)  & \cos(-\theta_1) \end{array} \right)
\left(\begin{array}{c} \nu_{e f} \cr -e^{i\xi}\nu_{\mu
f}\end{array}\right).
\label{eq:rephase}
\end{equation}
So doing, for all {\em exact} solutions, $K_\nu$  becomes unitary.

However, exact solutions are purely academic since, for example,
the absence of MCNC's is expected to be only approximate
\footnote{Note that (b) $\cup$ (f\raise-2pt\hbox{\tiny 2}) or (c) $\cup$
(g\raise-2pt\hbox{\tiny 2}), which entail
the exact absence of MCNC, is enough to have a single mixing angle,
since they also entail exact universality (a).}
.  As for exact universality (a), it is by itself not enough to have a
unique mixing angle since, in particular, (f\raise-2pt\hbox{\tiny 2}) or
(g\raise-2pt\hbox{\tiny 2}) may not be
satisfied. Moreover, (a)  may  be only approximately realized.
So, in the vicinity of the solutions above, a single mixing angle is not
enough to describe the system.

Another characteristic of the discrete solutions is their low sensitivity
to  small translations in the $(\theta_2,\theta_1)$ plane.
If one varies, for example, $\theta_2$, by  $\epsilon$ 
close to a specific point,  the l.h.s.'s of the universality condition,
$(c_2^2+s_1^2) -(c_1^2+s_2^2)=0$, and of the condition for the absence of
  MCNC's, $c_2s_2 \pm c_1s_1=0$, vary respectively
by $-4\epsilon\, c_2 s_2$ and $\epsilon(c_2^2 -s_2^2)$. Hence: 
\newline\null
\null\quad -\ at the discrete values $m\pi/2 + n\pi$, the
universality condition is satisfied at ${\cal O}(\epsilon^2)$
while the MCNC condition is only satisfied at ${\cal O}(\epsilon)$.
Referring to Fig.~2, this means that, if one varies
by $\epsilon$ the  angle (which is then a right angle)
between the two white dotted vectors, the (right) angle between
the blue and red vectors (and the one between
 the green mass eigenstates) also vary by $\epsilon$, while their
(unit) lengths are only altered at ${\cal O}(\epsilon^2)$;\newline
\null\quad -\ at the ``maximal mixing'' values, the reverse holds: the
absence of MCNC's is specially enforced; by the same variation as above,
it is now the angle between the red and blue vectors
(and the one between the mass eigenstates) which only varies by
$\epsilon^2$, while their lengths are
altered at  ${\cal O}(\epsilon)$; \newline \null
\quad -\ outside the set of discrete solutions, in particular for
Cabibbo-like solutions, both variations are instead ${\cal
O}(\epsilon)$.\newline
The absence of MCNC's is thus specially enforced at maximal mixing,
while universality is at angles $m\pi/2 + n\pi$.

As seen on Fig.~1, Cabibbo-like systems, characterized by a single
unconstrained mixing angle, can only be:\newline\null
\quad\ -  degenerate particles $z_1=z_2$
(in which case $\omega^2_1 = \varphi^2_m$ and $\omega^1_2 = \varphi^1_m$
such that $(\varphi^1_m,\varphi^2_m)$ form an orthonormal
basis);\newline\null
\quad\ -  ``off shell'' systems $(\psi^1(z), \psi^2(z))$ evaluated
at a common scale $z=q^2$, like quarks, for which the mixing angle is
$\theta(z)$;\newline\null
\quad\ -  very special systems like the ones satisfying  eqs.~(81)(82) of
 \cite{MaNoVi}.
\newline\null
Physical non-degenerate mesonic systems like $K^0-\overline{K^0}$
correspond to the other category (non Cabibbo-like):
when $CP$ (or exact family
symmetry) holds, mixing angles are identical and maximum, but when
$CP$ is broken, two angles occur, as shown in \cite{MaNoVi},
which are only close to maximum.
Such systems are expected to lie inside the small
(2-dimensional) areas in the vicinity of the discrete solutions (the
extended dots of Fig.~3), 
and not inside 1-dimensional deformations of exact Cabibbo-like systems, that
stay on the red or blue lines.

\subsubsection{Charged weak currents of mass eigenstates. Short
comments on  oscillations.}

Charged weak currents are coupled through $K_\nu^\dagger K_\ell$, the
so-called PMNS matrix \cite{PMNS}. 
Since charged leptons are non-degenerate coupled fermions too,
we expect, like previously obtained for neutrinos,
 the occurrence of a discrete set of mixing
angles $\pi/4\!\!\mod\!\pi/2$ and $\pi/2\!\!\mod\!\pi$.
$K_\ell$, like $K_\nu$, lies
accordingly  close to one of the ``academic'' unitary matrices evoked
above, such that $K_\nu^\dagger K_\ell$ should also be close to a
unitary matrix with a mixing angle in the same set of discrete values
\footnote{After convenient rephasing of the fermions (see
(\ref{eq:rephase})), both $K_\ell$ and
$K_\nu$ become close to unitary matrices with respective mixing angles
$\theta_\ell$ and $\theta_\nu$; the PMNS matrix is then close to a unitary
matrix with angle $(\theta_\ell -\theta_\nu)$.}
.
Several cases arise, the relevance of which with respect to oscillations we
would like to briefly discuss:

$\ast$\ if one among $K_\nu$ and $K_\ell$ is close to ``maximal''
and the other close to 
a multiple of $\pi/2$, the PMNS matrix is  close to ``maximal'';

$\ast$\ if both $K_\ell$ and $K_\nu$ are close to ``maximal'' with
respective mixing angles $(2k+1)\pi/4$ and $(2n+1)\pi/4$, the PMNS matrix
is close to a matrix with  mixing angle $(k-n)\pi/2$; this includes
the diagonal unit matrix (up to an irrelevant sign) and
the antidiagonal unit matrix;

$\ast$\ if both mixing angles of $K_\ell$ and $K_\nu$ are close to a
multiple of $\pi/2$, the same result holds.

Let us first stress that, while neutrino oscillations
are determined by $K_\nu$ alone, the detection of neutrinos on earth
always goes through their coupling to charged leptons,
which involves the PMNS matrix. We will consider two configurations for the
latter which both seem able to reproduce the observed solar electron neutrino
deficit on earth.

``Measuring'' a PMNS matrix close to maximal for two generations
favors the first possibility.  One among the two sets $(\nu_e, \nu_\mu)$
and  $(e^-, \mu^-)$ of leptons has then a maximal mixing, while the
mixing angle of the second is a multiple of $\pi/2$ (in which case only
simple mass-flavour alignment or nearly perfect ``crossing'' can occur).
The following picture may then be conceived.
Let us suppose that the flux of neutrinos stays
unperturbed during its travel from the center of the sun to the surface of
the earth, where it is detected. This can for example happen if the
Mikheyev-Smirnov-Wolfenstein (MSW)
effect \cite{MSW} does not operate inside the sun and if, then,
 vacuum oscillations
do not modify the neutrino spectrum. Its detection through the charged
currents (and, so, through the  maximal PMNS matrix) introduces a coefficient
$\approx \pm 1/\sqrt{2}$, which yields  a factor
$1/2$  in the square of the corresponding amplitude. A $1/2$
``deficit''  occurs though, in reality, no oscillation took
place.\newline\null
Mass-flavour alignment for one fermion species,
which is one of the two alternatives leading to
this first possibility, rules out the corresponding oscillations. It is 
natural to assume this property  for charged leptons (as usually done),
since such oscillations cannot anyhow be observed as soon as one measures
their energy  with a precision much higher than their mass-splitting
\cite{Kayser2}. The emerging picture may appear  coherent, though
the asymmetry arising between the two species of leptons raises
questions concerning the role of the electric charge.
Another possible weakness of this point of view lies
in the importance acquired by the measuring process through which,
furthermore, the
determination of the PMNS matrix cannot be truly asserted.

Now, we would like to point out that the following scenario, with
a PMNS matrix $\approx diag(1,1)$, is possible as well.
This belongs to the second possibility in the
original list, and accordingly provides a symmetric treatment
of neutral and charged leptons, which both have maximal mixing.
In this case, we are led instead to consider that neutrinos do
oscillate in their travel from the core of the sun to the earth. 
So, with respect to what is expected from solar models, a modified flux of
 $\nu_{em}$  reaches the earth, which can for example be altered by
a factor $\approx \pm 1/\sqrt{2}$.
These neutrinos then diagonally couple, in the detector, to charged
leptons with the coefficient $1$ occurring now in the PMNS matrix,
such that a global factor $1/2$ again occurs
in the (amplitude)\raise 4pt \hbox{\tiny 2}.
It can rightly be interpreted as ``neutrino oscillations''.

This treatment avoids the slightly opportunistic eviction
of electron-muon oscillations that occurred before, and which is not
mandatory:
it indeed undoubtedly leads to potential such oscillations, which can
appear problematic, but can be
argued away, as already explained, according to \cite{Kayser2}.
Charged currents  now differ from those in the quark sector by a
stronger suppression of their off-diagonal components as compared with the ones
obtained from the CKM matrix.

The two scenarii just described, which differ,
seem nevertheless to lead to the same conclusion, {\em i.e.} the observed
depletion of electronic solar neutrinos on earth. 
Discriminating between them (and others\;?) needs a more careful
investigation  which lies beyond the scope of this work.

As for the third possibility, it can easily be shown never to lead to any
neutrino deficit. Thus, the maximal character of $K_\nu$, which is 
common to the first two cases,  appears  as the essential
ingredient for the occurrence of this phenomenon.

\section{Neutral mesons}

Neutral kaons are composite states, and any Lagrangian that limits to their
description can only be effective. A similar propagator
 formalism can nevertheless be applied \cite{MaNoVi}\cite{Novikov}.

The second order electroweak  transitions,
which couple $K^0$ to $\overline{K^0}$,  are family changing transitions
in which $d$ and $s$ quarks get swapped.
When $CP$ is conserved, the $K^0_1$ and $K^0_2$ mass eigenstates,
respectively symmetric and antisymmetric with respect to $d
\leftrightarrow s$ family exchange, correspond to exact maximal mixing.
They form an orthonormal basis and no spurious state occurs.
$CP$ violation  alters this situation:
the $CP$ violating parameters $\epsilon_L$ and $\epsilon_S$ for 
$K_L$ and $K_S$ mass eigenstates 
slightly differ due to their  mass splitting and the Hamiltonian is no longer
normal.
Inside each {\em in} or {\em out} space mass eigenstates no longer
form orthonormal basis while {\em in} and {\em out} mass eigenstates, which
differ, form a bi-orthogonal basis.

The striking similarity between the latter and neutrinos suggests that the
symmetry by exchange of families (universality) plays an important role in
the nature of physical states.

Composite states (mesons) are however more complex that
fundamental particles. Indeed,
while the underlying electroweak theory for quarks does satisfy the
criteria of universality and absence of flavour / mass changing
neutral currents, the corresponding two types of conditions are not directly
available in an effective theory for neutral kaons alone.
Whether or not they could be implemented in a larger frame of an effective
theory for all scalar and pseudoscalar mesons, in which a general mass
matrix in flavour space should be diagonalized (see for example
\cite{Machet}), is a forthcoming matter of investigation.

\section{Conclusion}
\label{section:conclusion}

In this short note, we have proposed an enlargement of the mixing scheme
between mass and flavour eigenstates, which incorporates the 
peculiarity of both neutrinos and neutral kaons that their mixing angles
are close to maximal.
In continuation of \cite{MaNoVi}\cite{Novikov} we have shown that,
in quantum field theory, the mixing matrices of on-shell coupled
mass-split fermions  should
not be parametrized as unitary. The physics of two massive
neutrinos is then not that of a single mixing angle, but of two.
A new family of discrete mixing angles then springs out, among
which lies  the quasi-maximal mixing observed for neutrinos
and neutral kaons. When two different mixing angles are concerned, the
naive $\theta_2 \to \theta_1$ (``Cabibbo'') limit does not exist, which
explains how discrete solutions can easily be overlooked.
\newline\null
The role of  family exchange symmetry has been emphasized.
The generalization of this simple exercise
to more than two flavours will be the subject of a subsequent work.
Other aspects of coupled fermionic systems will also appear in
\cite{DuMaNoVi}.

\vskip .7cm

{\underline{\em Acknowledgments:}} Comments and suggestions by
 O. Babelon, A. Djouadi, O. Lychkovskiy, V.A. Novikov, L.B. Okun,
M.I. Vysotsky and J.B. Zuber  are gratefully acknowledged.
The authors bear of course full responsibility for their assertions. 

\newpage\null

\appendix

{\bf\Large Appendix}

\vskip 1cm

\section{The mass matrix in electroweak physics}
\label{section:mama}

The Yukawa couplings between fermions and the $SU(2)$ Higgs doublet
yield mass terms which 
write (we take the example of $u$-type quarks)
$\overline{\psi_{ufL}}M_u \psi_{ufR}$, to which are always added their
complex conjugate  $\overline{\psi_{ufR}}M_u^\dagger \psi_{ufL}$
(in order to  make the corresponding operator in the Lagrangian hermitian).
$\psi_{ufL}$ denotes the vector $(u_{fL},c_{fL},t_{fL})^T$ of
left-handed flavour eigenstates, $\psi_{ufR}$ its right counterpart.

The usual way  to make the CKM matrix appear in the weak bare Lagrangian
is to diagonalize the complex and potentially non-hermitian $M_u$
by a bi-unitary transformation $U_u^\dagger M_u V_u = D_u$.
One accordingly  {\em defines} quark masses as the square roots of
the (real positive) eigenvalues of the hermitian $M_uM_u^\dagger$ and
$M_u^\dagger M_u$. They  accordingly differ from the eigenvalues
of $M_u$ or $M_u^\dagger$ and they are real positive (thus presumably
not suitable to describe unstable fermions). The CKM
matrix, which acts on left-handed quarks, is $K= U_u^\dagger U_d$
(the $d$ subscript is attached to $d$-like quarks). The
mixing matrices $V_u$ and $V_d$ for right-handed quarks do not appear in
the bare electroweak Lagrangian (right handed quarks do not couple to weak
gauge bosons and  their vectorial couplings  to other gauge fields are,
 like  their kinetic terms, invariant by any unitary transformation).
Both initial mass term and its complex conjugate are
diagonalized by the same bi-unitary transformation, and
they   rewrite ($D_u$ being real)
$\overline{(U_u^\dagger \psi_{ufL})} D_u (V_u^\dagger \psi_{ufR})
+ \overline{(V_u^\dagger \psi_{ufR})} D_u (U_u^\dagger \psi_{ufL})$.
Two different mass bases then occur, $\psi_{umL} = U_u^\dagger \psi_{ufL}$
for left fermions and  $\chi_{umR} = V_u^\dagger \psi_{ufR}$
for right fermions, in terms of which the  mass term finally takes the
following form:
$\overline{\psi_{umL}} D_u \chi_{umR} + \overline{\chi_{umR}} D_u
\psi_{umL}$, which is that of a set of Dirac fermions built from the
two sets of Weyl fermions $\psi_{umL}$ and $\chi_{umR}$.
Kinetic terms can also be rewritten as the ones of these Dirac
fermions.

Another (equivalent) argument uses the {\em polar decomposition theorem}
stating that any complex matrix can always be written as the product of an
hermitian matrix times a unitary matrix $M_u = H_u {\cal W}_u$.
One then absorbs the unitary matrix ${\cal W}_u$ into the right-handed
fermions $\tilde\psi_{ufR} = {\cal W}_u \psi_{ufR}$, arguing again that
they do not couple
to the weak gauge bosons, and one is left with a hermitian mass matrix
\cite{Jarlskog2}
$H_u$ which can be diagonalized with a single unitary matrix $U_u$.
The mass term then becomes
$\overline{(U^\dagger_u \psi_{ufL})} D_u (U^\dagger_u
\tilde\psi_{ufR})$.  As already mentioned, the kinetic
terms for right-handed fermions, which are insensitive to any unitary
transformation, are in particular unaffected by their redefinition.
Like before, the  mass bases for right- and left-handed
fermions differ.

Thus, when the mass matrix is non-hermitian, two types of mixing
matrices have to be introduced for left and right fermions.  
The flavour content of the two spinors making up  mass eigenstates
is then different, unlike what occurs (by convention) for bispinor flavour
eigenstates.
If one rather sticks to identical transformations for left and right fermions,
the mass term $\overline{\psi_{ufL}}M_u \psi_{ufR} +
\overline{\psi_{ufR}}M_u^\dagger \psi_{ufL}$ generated by the coupling to
the Higgs doublet  includes a $\gamma^5$ term proportional to
$M_u-M_u^\dagger$  which must be taken into account.
A certain type of left-right symmetry is thus seen to be broken.

The necessary transformation of right fermions plays a role at
the quantum level (in the renormalized Lagrangian) because they  do couple
to weak gauge bosons when radiative corrections are included (at order
$g^4$) \cite{Djouadi}.

\section{Unitarity for an effective $\boldsymbol{CP}$-violating theory of
neutral kaons \cite{MaNoVi}}
\label{section:unit}

The following simple exercise can be done for neutral kaons, which proves that
the $K_L \to K_L$ transition amplitude stays unaltered by off-diagonal
transitions and the non-orthogonality of mass eigenstates. This happens,
as proved in \cite{MaNoVi}, when their two $CP$ violation parameters
are different $\epsilon_L \not= \epsilon_S$.

Consider first the simplest possible theory of an (uncoupled)
unique generic kaon $K$; the only possible transition is that of $K$ into
itself, with, of course, probability $1$. The propagator $\Delta$
 of $K$ is that of a free field.

Consider now the effective theory for the coupled $K^0-\overline{K^0}$ system
with $CP$ violation like in \cite{MaNoVi}. On one side,
since $K_L$ and $K_S$ are mass-split and no particle can carry away
the missing energy in a $K_L \to K_S$ on-shell decay,
the only open channel for the evolution of $K_L$ is itself.
On the other side, the Lagrangian includes a $K_L-K_S$
operator $\Big[ |K_S>_{in}{_{out}\!<K_L|} - |K_L>_{in}{_{out}\!\!<K_S|}\Big]$
with a coefficient $V(z)$ given in eq.~(114) of \cite{MaNoVi}, and
$K_L$ and $K_S$ are not orthogonal ${_{out}\!<K_S\ |\
K_L>_{in}} = -{_{out}\!<K_L\ |\ K_S>_{in}} = 
 \frac12\left(\frac{\xi_S}{\xi_L} - \frac{\xi_L}{\xi_S}\right)=\upsilon 
\not = 0,\ 
\xi_{L,S} = \sqrt{\frac{1-\epsilon_{L,S}^{in}}{1+\epsilon_{L,S}^{in}}}$.
\footnote{No confusion
should arise with asymptotic states of the S-matrix: {\em in} and {\em out}
states denote here the right and left eigenvectors of the 
kaon propagator, which is non-normal when $CP$ is broken.}

$A_{LL}$ and $A_{LS}$ (see Fig.~4) are the V-resummed diagonal
$K_L-K_L$ and mixed
 $K_L-K_S$ propagators; 
$A_{LL} ={\Delta}\left(\frac{1}{1+V^2\Delta^2}\right)$
and $A_{LS}= {\Delta}\left(\frac {-V\Delta}{1+V^2
\Delta^2}\right)$, where $\Delta \approx \Delta_L \approx \Delta_S$ is the
``average'' neutral kaon propagator.

The non-orthogonality of $K_L$ and $K_S$ entails that
any ingoing or outgoing $K_L$ has a non-zero $K_S$ component, and {\em vice
versa}. This brings additional corrections to $A_{LL}$ and $A_{LS}$, such
that the $K_L-K_L$ and $K_L-K_S$ full propagators are finally given by
$B_{LL}$ and $B_{LS}$ (we limited the expansion at ${\cal O}(\upsilon^2)$
for $B_{LL}$).


\vbox{
\begin{center}
\includegraphics[height=4truecm,width=12truecm]{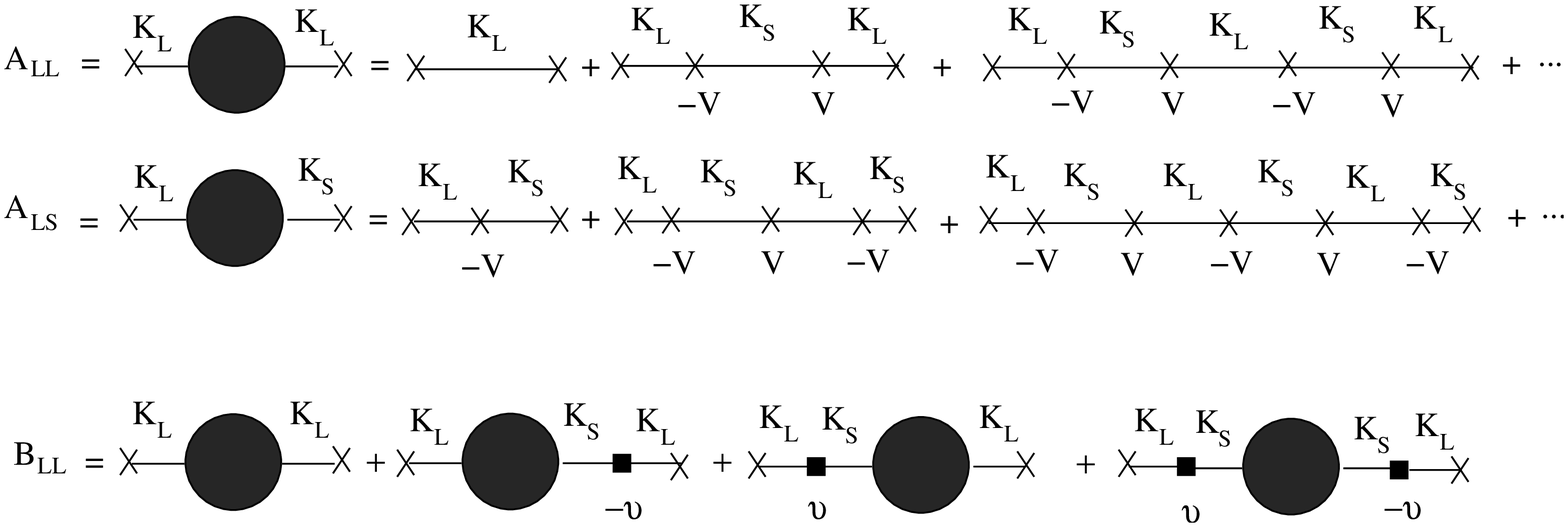}
\end{center}

\centerline{{\em Fig.~4: $K_L \to K_L$  transitions}}
}

\figskip

In the expression of $V(z)$ (eq.~(114) in \cite{MaNoVi}):\newline\null
-\  $b(z)$ and $c(z)$, which describe $K^0-\overline{K^0}$ transitions,
 are of second order $(g^4)$ in the weak interactions;\newline\null
-\  $e^{i\alpha}c(z)\stackrel{CP}{=} e^{-i\alpha}b(z)$
 (the corrections due to $CP$ violation are proportional to
$(\epsilon_S^{in}+\epsilon_L^{in})(e^{i\alpha}b(z) + e^{-i\alpha}c(z))$);
\newline\null
-\ $D(z) = 1 + {\cal O}(\epsilon^2)$, $a(z) \approx \frac{1}{\Delta(z)}$;
\newline\null
-\  the coefficient which factorizes $a(z)$ is the same $2\upsilon$ as  above.
\newline\null
Since $a(z)$ in the diagonal part of the inverse propagator $a(z) \approx
\frac{1}{\Delta(z)}$  and since
one can approximate  $V(z)\approx \frac{1}{\Delta(z)}\, \upsilon$, one gets
$A_{LL}\approx {\Delta}\left(\frac{1}{1+\upsilon^2}\right)$
and $A_{LS}\approx-{\Delta}\left(\frac {\upsilon}
{1+\upsilon^2 }\right)$; likewise, one finds  $A_{SS} \approx A_{LL}$ and
$A_{SL} \approx -A_{LS}$, which finally gives $B_{LL} \approx {\Delta}$:
it is the ``free'' kaon propagator, like in a theory where
 none of the intricacies due to mass splitting and $CP$ violation (in
particular the non-unitarity of the mixing matrix) occurs. The 
transition probability for $K_L \to K_L$ thus  stays also unchanged
and equal to $1$: no violation of unitarity occurs.
If neutral kaons
are given complex $(mass)^2$ to account for their instability, the
probability is no longer conserved, but this violation of unitarity
is unrelated with the peculiarities of the mixing matrix.

%
\newpage\null

\begin{em}

\end{em}


\begin{thebibliography}{50}

\bibitem{Kayser}
See for example:\newline\null
B. KAYSER: ``Neutrino Physics'' hep-ph/0506165, and references therein.

\bibitem{AFM}
See for example:\newline\null
G. ALTARELLI, F. FERUGLIO \& I. MASINA: hep-ph/0402155, Nucl. Phys. B 689
(2004) 157-171, and references therein.

\bibitem{Ma}
E. MA: ``Lepton Family Symmetry and the Neutrino Mixing Matrix'',
hep-ph/0606024, and references therein.

\bibitem{Alt}
G. ALTARELLI: ``Normal and Special Models of Neutrino Masses and Mixings'',
 hep-ph/0508053, and references therein.

\bibitem{Smir}
A.Yu. SMIRNOV: ``Neutrino masses and mixing: leptons versus quarks''
hep-ph/0604213, and references therein.

\bibitem{Jarlskog1}
C. JARLSKOG: Phys.Lett. B 625 (2005) 63-66;\newline\null
ibidem: ``The Role of Invariant Functions in Understanding Masses and
Mixings'', hep-ph/0606050.

\bibitem{MaNoVi}
B. MACHET, V.A. NOVIKOV \& M.I. VYSOTSKY:
hep-ph/0407268, Int. J. Mod. Phys. A 20 (2005) 5399-5452,
and references therein.

\bibitem{Novikov}
V.A. NOVIKOV: ``Binary systems in QM and in QFT: CPT'', hep-ph/0509126.

\bibitem{PMNS}
B. PONTECORVO: Sov. Phys. JETP 26 (1968) 984 [Zh. Eksp. Teor. Fiz. 53
(1967) 1717];

Z. MAKI, M. NAKAGAWA \& S. SAKATA: Prog. Theor. Phys. 28 (1962) 870.

\bibitem{Kayser2}
B. KAYSER: Phys. Rev. D 24 (1981) 110.

\bibitem{MSW}
L. WOLFENSTEIN: Phys. Rev. D 17 (1978) 2369;\newline\null
 ibidem: Phys. Rev. D 20 (1979) 2634;

S.P. MIKHEYEV \& A.Yu. SMIRNOV: Yad. Fiz. 42 (1985) 1441 [Sov. J. Nucl.
Phys. 42 (1985) 913];\newline\null
ibidem: Il Nuovo Cimento C 9 (1986) 17;\newline\null
ibidem: Zh. Eksp. Teor. Fiz. 91 (1986) 7 [Sov. Phys. JETP 64 (1986) 4].

\bibitem{Machet}
B. MACHET: Phys. Lett. B 385 (1996) 198-208.

\bibitem{DuMaNoVi}
Q. DURET, B. MACHET, V.A. NOVIKOV \& M.I. VYSOTSKY: ``Fermions in Quantum
Field Theory'', in preparation.

\bibitem{Jarlskog2}
See for example:\newline\null
C. JARLSKOG: Phys. Rev. Lett. 55 (1985) 1039.

\bibitem{Djouadi}
A. DJOUADI: private communication.

\end{thebibliography}
\end{document}